\documentclass[twocolumn,showpacs,amsmath,amssymb,prl,superscriptaddress,floatfix]{revtex4}
\usepackage{graphicx}
\usepackage{dcolumn}
\usepackage{bm}
\begin{document}
\title{Momentum dependence of the $\boldsymbol{\phi}$-meson nuclear transparency}
\thanks{Based in part on a PhD thesis submitted by one of the authors (AP) to ITEP, Moscow.}
\author{M.~Hartmann}\email[E-mail: ]{m.hartmann@fz-juelich.de}%
\affiliation{Institut f\"ur Kernphysik and J\"ulich Centre for Hadron
Physics, Forschungszentrum J\"ulich, D-52425 J\"ulich, Germany}
\author{Yu.T.~Kiselev}\email[E-mail: ]{yurikis@itep.ru}%
\affiliation{Institute for Theoretical and Experimental Physics,
RU-117218 Moscow, Russia}
\author{A.~Polyanskiy}%
\affiliation{Institut f\"ur Kernphysik and J\"ulich Centre for Hadron
Physics, Forschungszentrum J\"ulich, D-52425 J\"ulich, Germany}
\affiliation{Institute for Theoretical and Experimental Physics,
RU-117218 Moscow, Russia}
\author{E.Ya.~Paryev}
\affiliation{Institute for Nuclear Research, Russian Academy of
Sciences, RU-117312 Moscow, Russia}
\author{M.~B\"uscher}
\affiliation{Institut f\"ur Kernphysik and J\"ulich Centre for Hadron
Physics, Forschungszentrum J\"ulich, D-52425 J\"ulich, Germany}
\author{D.~Chiladze}
\affiliation{Institut f\"ur Kernphysik and J\"ulich Centre for Hadron
Physics, Forschungszentrum J\"ulich, D-52425 J\"ulich, Germany}
\affiliation{High Energy Physics Institute, Tbilisi State University, GE-0186
Tbilisi, Georgia}
\author{S.~Dymov}
\affiliation{Physikalisches Institut, Universit{\"a}t
Erlangen-N\"urnberg, D-91058 Erlangen, Germany}
\affiliation{Laboratory of Nuclear Problems, Joint Institute for
Nuclear Research, RU-141980 Dubna, Russia}
\author{A.~Dzyuba}
\affiliation{High Energy Physics Department, Petersburg Nuclear
Physics Institute, RU-188350 Gatchina, Russia}
\author{R.~Gebel}
\affiliation{Institut f\"ur Kernphysik and J\"ulich Centre for Hadron
Physics, Forschungszentrum J\"ulich, D-52425 J\"ulich, Germany}
\author{V.~Hejny}
\affiliation{Institut f\"ur Kernphysik and J\"ulich Centre for Hadron
Physics, Forschungszentrum J\"ulich, D-52425 J\"ulich, Germany}
\author{B.~K\"{a}mpfer}
\affiliation{Institut f\"ur Kern- und Hadronenphysik,
Helmholtz-Zentrum Dresden-Rossendorf, D-01314 Dresden, Germany}
\author{I.~Keshelashvili}
\affiliation{Department of Physics, University of Basel, CH-4056 Basel,
Switzerland}
\author{V.~Koptev}\thanks{Deceased}
\affiliation{High Energy Physics Department, Petersburg Nuclear
Physics Institute, RU-188350 Gatchina, Russia}
\author{B.~Lorentz}
\affiliation{Institut f\"ur Kernphysik and J\"ulich Centre for Hadron
Physics, Forschungszentrum J\"ulich, D-52425 J\"ulich, Germany}
\author{Y.~Maeda}
\affiliation{Research Center for Nuclear Physics, Osaka
University, Ibaraki, Osaka 567-0047, Japan}
\author{V.~K.~Magas}
\affiliation{Departament d'Estructura i Constituents de la Mat\`eria and
Institut de Ci\`encies del Cosmos, Universitat de Barcelona, E-08028
Barcelona, Spain}
\author{S.~Merzliakov}
\affiliation{Institut f\"ur Kernphysik and J\"ulich Centre for Hadron
Physics, Forschungszentrum J\"ulich, D-52425 J\"ulich, Germany}
\affiliation{Laboratory of Nuclear Problems, Joint Institute for
Nuclear Research, RU-141980 Dubna, Russia}
\author{S.~Mikirtytchiants}
\affiliation{Institut f\"ur Kernphysik and J\"ulich Centre for Hadron
Physics, Forschungszentrum J\"ulich, D-52425 J\"ulich, Germany}
\affiliation{High Energy Physics Department, Petersburg Nuclear
Physics Institute, RU-188350 Gatchina, Russia}
\author{M.~Nekipelov}
\affiliation{Institut f\"ur Kernphysik and J\"ulich Centre for Hadron
Physics, Forschungszentrum J\"ulich, D-52425 J\"ulich, Germany}
\author{H.~Ohm}
\affiliation{Institut f\"ur Kernphysik and J\"ulich Centre for Hadron
Physics, Forschungszentrum J\"ulich, D-52425 J\"ulich, Germany}
\author{L.~Roca}
\affiliation{Departamento de F\'isica, Universidad de Murcia, E-30071 Murcia, Spain}
\author{H.~Schade}
\affiliation{Institut f\"ur Kern- und Hadronenphysik,
Helmholtz-Zentrum Dresden-Rossendorf, D-01314 Dresden, Germany}
\affiliation{Institut f\"ur Theoretische Physik, TU Dresden, D-01062 Dresden, Germany}
\author{V.~Serdyuk}
\affiliation{Institut f\"ur Kernphysik and J\"ulich Centre for Hadron
Physics, Forschungszentrum J\"ulich, D-52425 J\"ulich, Germany}
\affiliation{Laboratory of Nuclear Problems, Joint Institute for
Nuclear Research, RU-141980 Dubna, Russia}
\author{A.~Sibirtsev}
\affiliation{Physikalisches Institut, Universit{\"a}t Erlangen-N\"urnberg,
D-91058 Erlangen, Germany}
\author{V.~Y.~Sinitsyna}
\affiliation{P.~N.~Lebedev Physical Institute, RU-119991 Moscow, Russia}
\author{H.~J.~Stein}
\affiliation{Institut f\"ur Kernphysik and J\"ulich Centre for Hadron
Physics, Forschungszentrum J\"ulich, D-52425 J\"ulich, Germany}
\author{H.~Str\"oher}
\affiliation{Institut f\"ur Kernphysik and J\"ulich Centre for Hadron
Physics, Forschungszentrum J\"ulich, D-52425 J\"ulich, Germany}
\author{S.~Trusov}
\affiliation{Institut f\"ur Kern- und Hadronenphysik,
Helmholtz-Zentrum Dresden-Rossendorf, D-01314 Dresden, Germany}
\affiliation{Skobeltsyn Institute of Nuclear Physics, Lomonosov Moscow
State University, RU-119991 Moscow, Russia}
\author{Yu.~Valdau}
\affiliation{Institut f\"ur Kernphysik and J\"ulich Centre for Hadron
Physics, Forschungszentrum J\"ulich, D-52425 J\"ulich, Germany}
\affiliation{Helmholtz-Institut f\"ur Strahlen- und Kernphysik, Universit\"at
Bonn, D-53115 Bonn, Germany}
\author{C.~Wilkin}
\affiliation{Physics and Astronomy Department, UCL, London WC1E 6BT,
United Kingdom}
\author{P.~W\"ustner}
\affiliation{Zentralinstitut f\"ur Elektronik,
Forschungszentrum J\"ulich, D-52425 J\"ulich, Germany}
\author{Q.J.~Ye}
\affiliation{Institut f\"ur Kernphysik and J\"ulich Centre for Hadron
Physics, Forschungszentrum J\"ulich, D-52425 J\"ulich, Germany}
\affiliation{Physics Department, Duke University, Durham, NC 27708, USA}

\date{\today}

\vspace*{1cm}
\begin{abstract}
The production of $\phi$ mesons in proton collisions with C, Cu, Ag, and Au
targets has been studied via the $\phi\to K^+K^-$ decay at an incident beam
energy of 2.83~GeV using the ANKE detector system at COSY. For the first
time, the momentum dependence of the nuclear transparency ratio, the
in-medium $\phi$ width, and the differential cross section for $\phi$ meson
production at forward angles have been determined for these targets over the
momentum range of 0.6\,-\,1.6~GeV/$c$. There are indications of a significant
momentum dependence in the value of the extracted $\phi$ width, which
corresponds to an effective ${\phi}N$ absorption cross section in the range
of 14\,-\,21~mb.
\end{abstract}

\maketitle

\section{Introduction}

The study of the effective masses and widths of light vector mesons $\rho$,
$\omega$, $\phi$ in a nuclear medium, through their production and decay in
the collisions of photon, hadron and heavy--ion beams with nuclear targets,
has received considerable attention in recent years (see, for
example,~\cite{Rapp:2000,Hayano:2010,Leupold:2010}). This interest in the
in-medium properties was triggered by the hypothesis of universal scaling of
Brown and Rho~\cite{Brown:1991}, as well as by the QCD-sum-rule-based
prediction of Hatsuda and Lee~\cite{Hatsuda:1992} that the masses of these
mesons should be lower in nuclear matter due to the partial restoration of
chiral symmetry in hot and dense nuclear matter. This is a fundamental
symmetry of QCD in the limit of vanishing quark masses and its restoration
would be characterized by a reduction of the scalar quark condensate in the
medium compared to its magnitude in vacuum.

The natural width of the $\phi(1020)$ meson is only 4.3~MeV/$c^2$, which is
narrow compared to other nearby resonances. It is therefore the optimal probe
for the investigation of medium modifications because small effects should be
experimentally observable. However, this meson is of interest for other
reasons. Since the $\phi$ is an almost pure $s\bar{s}$ state, the
Okubo-Zweig-Iizuka (OZI) rule~\cite{OZI} suppresses quark exchange in its
interaction with ordinary (non-strange) baryonic matter. Gluon exchange,
which plays a substantial role in high-energy interactions between hadrons,
is therefore expected to dominate ${\phi}N$ scattering at all
energies~\cite{Mibe:2007}. The $\phi$ might therefore be considered as a
\textit{clean} system for the study of gluonic exchange which, at low
energies, should manifest itself as an attractive QCD van der Waals force,
which could even lead to the formation of ${\phi}N$ bound
states~\cite{Gao:2001}. Secondly, owing to the small energy release in the
dominant $K\bar{K}$ decay channel, any changes in the $\phi$ properties are
very sensitive to possible in-medium modifications of kaons and antikaons, a
subject which is also of great current interest.

The main modification of the $\phi$ properties in nuclear matter is expected
to be a broadening of its spectral function, while its mass should be hardly
shifted~\cite{Leupold:2010,Hatsuda:1992,Oset:2001,Klingl:1998,Cabrera:2002,Cabrera:2003}.
Dileptons from $\phi \to e^{+}e^{-}/\mu^{+} \mu^{-}$ decays experience no
strong final-state interactions in a nucleus. Any broadening of the $\phi$
line-shape in the nuclear matter should be directly testable by examining
dilepton mass spectra produced by elementary or heavy-ion beams, provided
that the necessary cuts are applied at low $\phi$
momenta~\cite{Muto:2005,Paryev:2005}. However, the measurement of such
spectra is difficult due to the low branching ratios for leptonic decays.
Furthermore, their sensitivity to in-medium modifications will be reduced by
the profile of the nucleus spanning all densities from zero to normal nuclear
matter density $\rho_0=0.16~\text{fm}^{-3}$, which smears out any
density-dependent signal~\cite{Leupold:2010}. The KEK-PS-E325 collaboration
measured $e^{+}e^{-}$ invariant mass distributions in the $\phi$ region in
proton-induced reactions on carbon and copper at 12~GeV and reported a mass
shift of 3.4\% and a width increase by a factor of 3.6 at density $\rho_0$
for $\phi$ momenta around 1~GeV/$c$~\cite{Muto:2005}.

An alternative way to determine the in-medium broadening of the $\phi$ meson
has been proposed in~\cite{Cabrera:2003} and adopted
in~\cite{Ishikawa:2004,Wood:2010}. Here, the variation of the $\phi$
production cross section (or nuclear transparency ratio) with atomic number
$A$ has been studied both experimentally and theoretically. The big advantage
of this method is that one can exploit the dominant $K^+K^-$ branching ratio
($\approx$ 50$\%$). The $A$-variation depends on the attenuation of the
$\phi$ flux in the nucleus which, in turn, is governed by the imaginary part
of the $\phi$ in-medium self-energy or width. In the low-density
approximation~\cite{Dover:1971}, this width can be related to an effective
$\phi N$ absorption cross section $\sigma_{{\phi}N}$, though this is less
obvious at higher densities where two-nucleon effects might be significant.

A large in-medium $\phi N$ absorption cross section of about 35~mb was
inferred in a Glauber-type analysis by the LEPS collaboration from
measurements of $K^+K^-$ pairs photoproduced on Li, C, Al and Cu targets at
SPring-8 for average $\phi$ momenta $\approx 1.8$~GeV/c~\cite{Ishikawa:2004}.
Similar considerations were given in~\cite{Sibirtsev:2006}. This large $\phi
N$ cross section was confirmed in BUU transport model
calculations~\cite{Muehlich:2006}. In the low-density approximation, this
implies an in-medium $\phi$ width of about 110~MeV/$c^2$ in its rest frame at
density $\rho_0$ for the conditions of the KEK measurements~\cite{Muto:2005}.
This is clearly incompatible with the width reported in the KEK experiment.
This value of $\sigma_{\phi N}$ is also significantly larger than the
$\approx 10$~mb obtained from both photoproduction data on hydrogen, using
the vector-meson dominance model in the photon energy range $E_{\gamma} <
10$~GeV~\cite{Behrend:1978,Sibirtsev:2006}, and the additive quark
model~\cite{Lipkin:1966}.

Values of the in-medium $\sigma_{\phi N}$ were also determined by the CLAS
collaboration from transparency ratio measurements at JLab~\cite{Wood:2010}.
In this experiment, the $\phi$ mesons were photoproduced on $^{2}$H, C, Ti,
Fe, Pb targets and detected via their $e^+e^-$ decay mode. From an analysis
of the transparency ratios normalized to carbon within a Glauber model,
values of $\sigma_{\phi N}$ in the range of 16\,-\,70~mb were extracted for
an average $\phi$ momentum of $\approx 2$~GeV/c. These are to be compared to
the free value of $\approx 10$~mb.

The SPring-8~\cite{Ishikawa:2004} and JLab~\cite{Wood:2010} results are
consistent with the JLab measurements of coherent~\cite{Mibe:2007} and
incoherent~\cite{Qian:2009} $\phi$ photoproduction from deuterium. The
coherent data suggest that $\sigma_{\phi N}\approx 30$~mb together with a
larger slope for elastic $\phi N$ scattering compared to ${\gamma}N \to
{\phi}N$~\cite{Mibe:2007}. A combined analysis of coherent and incoherent
$\phi$ meson photoproduction from deuterium favors $\sigma_{\phi N}
> 20$~mb~\cite{Qian:2009}.

Whereas medium modifications of $\sigma_{\phi N}$ might offer a plausible
explanation of the SPring-8~\cite{Ishikawa:2004} and JLab~\cite{Wood:2010}
data, they can hardly account for the large value found in deuterium. Nuclear
density effects are here minimal and other mechanisms, beyond medium
modifications, could be more important~\cite{Qian:2009}. In this context it
should be noted that the LEPS collaboration recently studied incoherent
$\phi$ photoproduction from deuterium at forward angles for
$E_{\gamma}=1.5$\,-\,2.4~GeV~\cite{Chang:2010}. The nuclear transparency
ratio, extracted as a function of $E_{\gamma}$, shows a significant
25\,-\,30\% reduction, which is consistent with that previously deduced by
the same collaboration from the $A$-dependence of this ratio for heavier
nuclear targets~\cite{Ishikawa:2004}. On the other hand, very recent data on
incoherent $\phi$ photoproduction on deuterium, taken by CLAS collaboration
in a similar photon energy range but over a region of larger momentum
transfers~\cite{Qian:2011}, are inconsistent with the LEPS
results~\cite{Chang:2010}.

The divergent conclusions drawn from the various experiments emphasize the
need to improve our understanding of the ${\phi}N$ interaction in nuclei.
With the aim of furthering these studies, we have measured the inclusive
production of $\phi$ mesons at forward angles in the collisions 2.83~GeV
protons with C, Cu, Ag, and Au targets. The meson was detected via the
$\phi\to K^+K^-$ decay using the ANKE-COSY magnetic
spectrometer~\cite{Polyanskiy:2011}. Values of the nuclear transparency ratio
normalized to carbon, $R=(12/A)(\sigma^A/\sigma^C)$,  were deduced, averaged
over the $\phi$ momentum range 0.6\,-\,1.6~GeV/$c$. Here $\sigma^A$ and
$\sigma^C$ are inclusive cross sections for $\phi$ production in $pA$
($A=$~Cu, Ag, Au) and $p$C collisions in the angular cone $\theta_{\phi} <
8^{\circ}$. The comparison of the ratio with model
calculations~\cite{Magas:2004,Paryev:2009,Rossendorf} yielded an in-medium
$\phi$ width of 33\,-\,50~MeV/$c^2$ in the nuclear rest frame for an average
$\phi$ momentum of 1.1~GeV/$c$ for normal nuclear density $\rho_0$.

Because of the large number of reconstructed $\phi$ mesons for each target
(7000\,-\,10000), the data could be put in bins in order to obtain
differential distributions. This allows us to carry out more detailed
investigations, in particular of the momentum dependence of the parameters.
In this paper we report on the results of further analysis of the data
collected in our experiment~\cite{Polyanskiy:2011,Hartmann:2010}.
%
%
\section{Experiment and Results}

A series of thin and narrow C, Cu, Ag, and Au targets was inserted in a beam
of 2.83~GeV protons, circulating in the COSY Cooler Synchrotron/storage ring
of the Forschungszentrum J\"ulich, in front of the main spectrometer magnet
D2 of the ANKE system (see~\cite{Barsov:2001,Hartmann:2007}). The ANKE
spectrometer has detection systems placed to the right and left of the beam
to register positively and negatively charged ejectiles which, in the case of
$\phi$ meson production, are the $K^+$ and $K^-$. Although only used here for
efficiency studies, forward-going charged particles could also be measured in
coincidence. The positively charged kaon was first selected using a dedicated
detection system that can identify a $K^+$ against a pion/proton background
that is $10^5$ more intense (compare Fig.~2 in~\cite{Hartmann:2010}
and~\cite{Hartmann:2007,Buescher:2002}). The coincident $K^-$ was
subsequently identified from the time-of-flight difference between the stop
counters in the negative and positive detector systems.

\begin{figure*}[ht]
  \vspace*{+0mm}
  \includegraphics[clip,width=1.6\columnwidth]{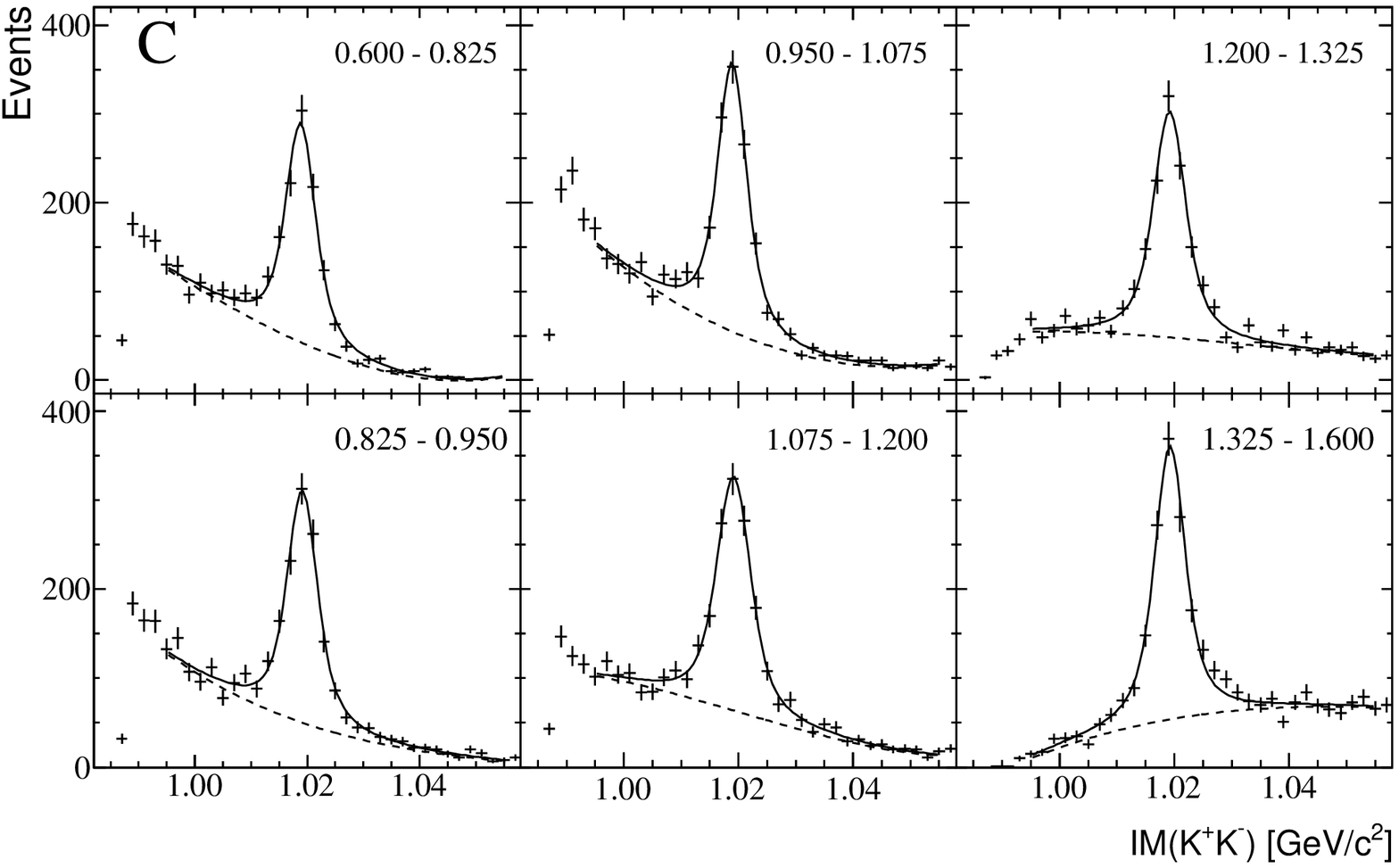}
  \includegraphics[clip,width=1.6\columnwidth]{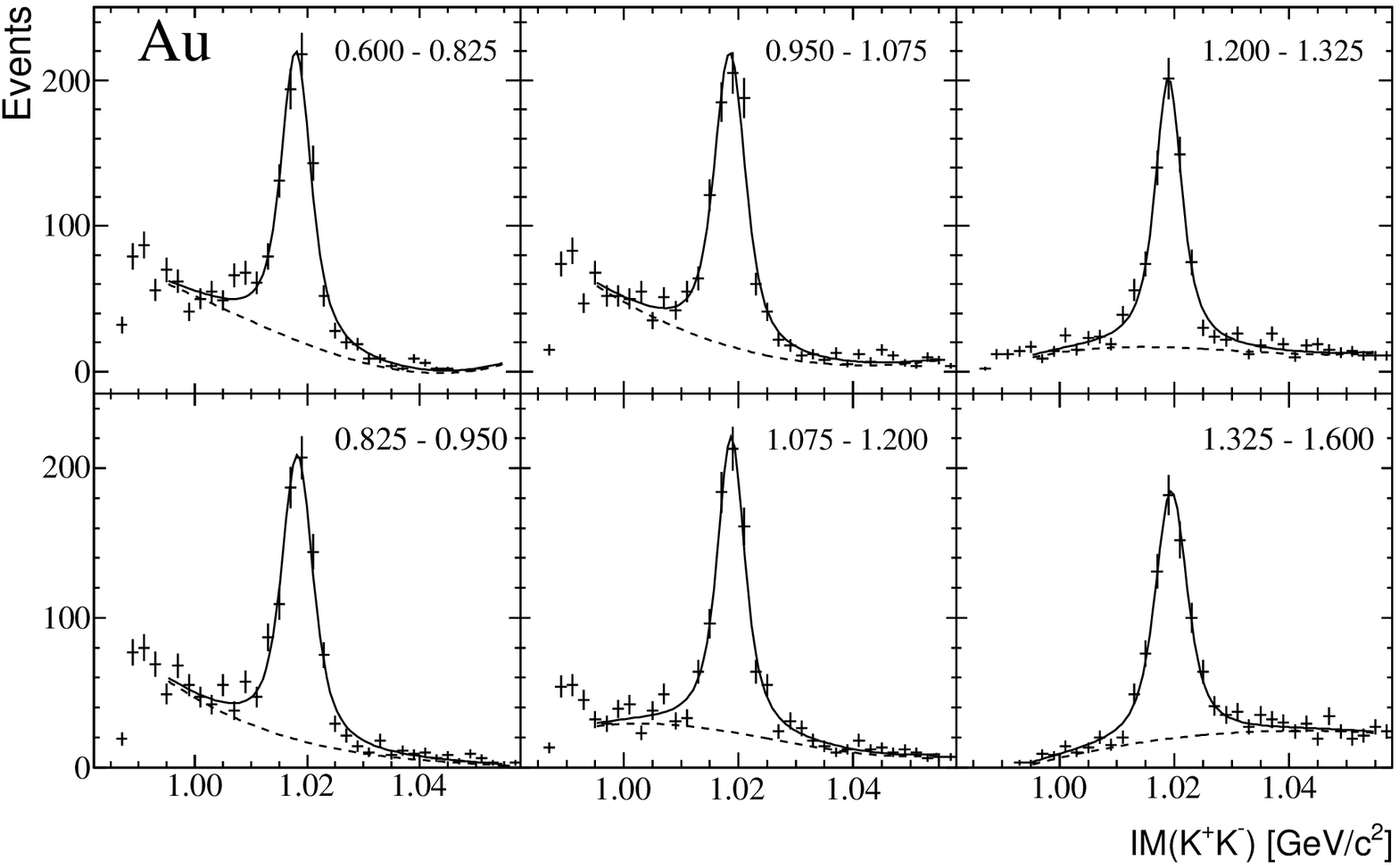}
  \vspace*{-2mm}
\caption{Invariant mass distributions of $K^{+}K^{-}$ pairs
produced in $p\,$C and $p$Au collisions in the $\phi$ momentum
bin noted in GeV/$c$. Fits to the uncorrected experimental data
in terms of an expected $\phi$ shape and a physical background
are shown by the solid lines. The dashed lines are third order
polynomial parameterizations of the backgrounds in the region
of the $\phi$ peak.} \label{fig:IM} \vspace*{-2mm}
\end{figure*}

The accessible range of the $\phi$ meson momenta, $0.6 < p_{\phi} <
1.6$~GeV/$c$, was divided into six intervals with about 1000 mesons per bin.
The $K^+K^-$ invariant mass spectra measured in the $pA \to K^+K^-X$ reaction
look similar for all four targets and only the results for carbon and gold
are presented in Fig.~\ref{fig:IM}. In every case there is a clear $\phi$
signal sitting on a background of mainly non-resonant kaon pair production
together with a relatively small number of misidentified events. To study the
momentum dependence of the transparency ratio $R$, the numbers of $\phi$
events falling within each momentum bin were evaluated for the four targets.
With this in mind, the mass spectra were fitted by the incoherent sum of a
Breit-Wigner function with the natural $\phi$ width, convoluted with a
Gaussian resolution function with $\sigma\approx 1$~MeV/$c^2$, and a
polynomial background function.

\begin{figure}[ht]
  \vspace*{+0mm}
  \includegraphics[clip,width=0.9\columnwidth]{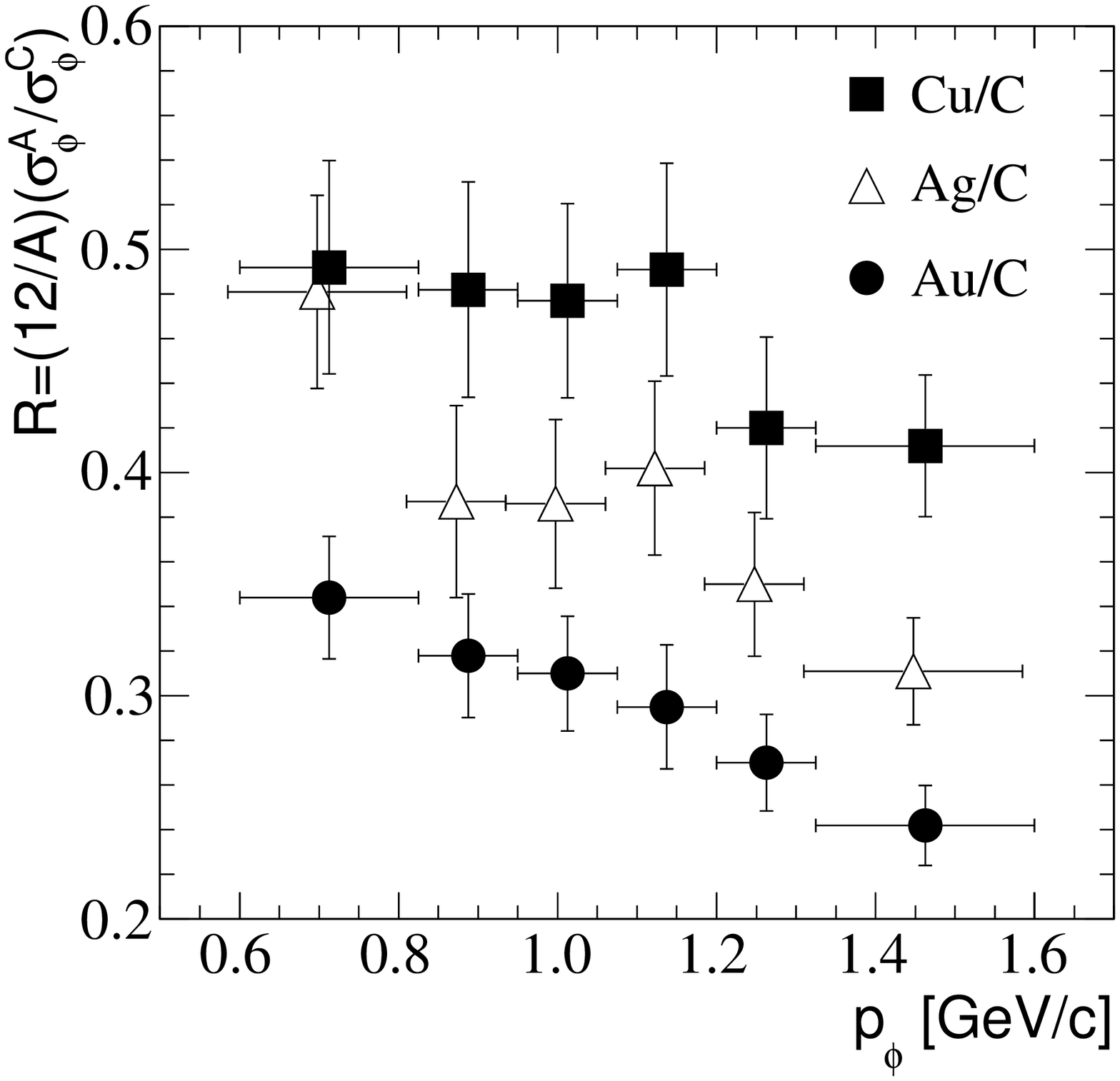}
  \vspace*{-2mm}
\caption{Momentum dependence of the transparency ratio $R$,
normalized to carbon, for Cu, Ag, and Au targets.
}\label{fig:TRExp} \vspace*{-2mm}
\end{figure}

Since the efficiency corrections in the ANKE spectrometer are essentially
target-independent, after taking the luminosity into account, to a good
approximation the ratio of the number of reconstructed $\phi$ in any bin for
a nucleus $A$ to that for carbon corresponds to ratio of the cross sections
for $\phi$ production on these targets in the given momentum
bin~\cite{Polyanskiy:2011}. The resulting transparency ratios are given in
Table~\ref{tab::table1} and shown in Fig.~\ref{fig:TRExp}. For all the
combinations, Cu/C, Ag/C and Au/C, $R$ falls when the $\phi$ momentum
increases.

\begin{table*}[ht]
\caption{The measured transparency ratio $R$ in the acceptance window of the
ANKE spectrometer for six momentum bins. The first errors are statistical and
the second systematic, which are mainly associated with the fit quality. In
addition there are overall systematic uncertainties in the ratios of about
5\,-\,6\%, coming principally from the relative
normalizations.\label{tab::table1}}
\begin{center}
\begin{tabular}{|c|c|c|c|}
\hline%
$p_{\phi}$ [GeV/$c$] & $R$(Cu/C) & $R$(Ag/C) & $R$(Au/C) \\
\hline
0.600--0.825 & $0.49\pm0.03\pm0.03$ & $0.48\pm0.03\pm0.03$ & $0.34\pm0.02\pm0.02$ \\
0.825--0.950 & $0.48\pm0.03\pm0.04$ & $0.39\pm0.03\pm0.03$ & $0.32\pm0.02\pm0.02$ \\
0.950--1.075 & $0.48\pm0.03\pm0.03$ & $0.39\pm0.03\pm0.03$ & $0.31\pm0.02\pm0.02$ \\
1.075--1.200 & $0.49\pm0.03\pm0.04$ & $0.40\pm0.03\pm0.03$ & $0.30\pm0.02\pm0.02$ \\
1.200--1.325 & $0.42\pm0.03\pm0.03$ & $0.35\pm0.02\pm0.02$ & $0.27\pm0.02\pm0.01$ \\
1.325--1.600 & $0.41\pm0.02\pm0.02$ & $0.31\pm0.02\pm0.02$ & $0.24\pm0.01\pm0.01$ \\
\hline
\end{tabular}
\end{center}
\end{table*}

Each momentum bin contains roughly equal numbers of events and the associated
statistical uncertainty in $R$ is about 6\,-\,7\%. The main systematic
effects arise from the evaluation of the number of $\phi$ falling within a
certain momentum bin and the overall uncertainty in the relative
normalization. The first was estimated by varying the fit parameters, the
binning of the histograms, the range of fitting, and the order of the
polynomial background. These results are reported in Table~\ref{tab::table1}.
The relative normalization uncertainty, which is described in detail
in~\cite{Polyanskiy:2011,Hartmann:2010}, is about 4\,-\,6\%, depending on the
target nucleus.

In order to test further the model calculations, the double differential
cross sections for $\phi$ meson production have been evaluated within the
ANKE acceptance window for each momentum bin $\Delta p$ and each nucleus $A$
as
\begin{equation}
\label{eq:CrSecPhi}
 \frac {d^{2} \sigma_{\phi}^A} {dp d\Omega}	
= \frac{1}{(\Delta p \Delta \Omega)}
\frac{N^{A}_{\phi}}{\langle \varepsilon_{\phi} \rangle L^{A}_{\textrm{int}}},
\end{equation}
where $N^A_{\phi}$ is the number of $\phi$ detected in a solid angle $\Delta
\Omega$ and $L^{A}_{\textrm{int}}$ is the integrated luminosity for target
$A$. In order to estimate the average efficiency for $\phi$ identification
$\langle \varepsilon_{\phi} \rangle$, the detection efficiency was first
evaluated for each nucleus and each momentum bin. For this the ratio of the
number of $\phi$ detected to that determined from the fitting the $K^+K^-$
efficiency-corrected invariant-mass distributions was calculated on an
event-by-event basis. These efficiencies were then averaged over the target
nuclei for each momentum bin. The root mean square deviations of the
individual efficiencies from $\langle \varepsilon_{\phi} \rangle$ were about
5\%, which is consistent with the statistical uncertainties.

The efficiency was estimated for each event using
\begin{equation}
\label{eq:EffPhi}
\varepsilon_{\phi} = \varepsilon_{\mathrm{tel}} \times
\varepsilon_{\mathrm{tr}} \times
\varepsilon_{\mathrm{acc}}  \times
\varepsilon_{\mathrm{BR}}.
\end{equation}
The track reconstruction efficiency of $K^+K^-$ pairs
$\varepsilon_{\textrm{tr}}$ is determined from the experimental data. The
correction for kaon decay in flight and acceptance
($\varepsilon_{\textrm{acc}}$) is determined as a function of the laboratory
momentum and the laboratory polar angle of the $\phi$ meson, using
simulations and assuming an isotropic $\phi$ decay in its rest frame. The
range-telescope efficiency $\varepsilon_{\textrm{tel}}$ is extracted from
calibration data on $K^+p$ coincidences. Finally, $\varepsilon_{\textrm{BR}}$
represents the branching ratio of the $\phi \to K^+K^-$ decay mode.

The integrated luminosity $L^{A}_{\textrm{int}}$ is calculated using the
measured flux of $\pi^+$ mesons with momenta $\approx 500$~MeV/$c$ produced
at small laboratory angles. Values of the $\pi^+$ production cross sections
used at 2.83~GeV are $59.8 \pm 7.2$ for carbon, $113 \pm 15$ for copper, $138
\pm 19$ for silver and $174 \pm 24$~mb/(sr\ GeV/$c$) for gold (the details
are given in~{\cite{Polyanskiy:Hadron2011}).

The measured double differential cross section for $\phi$ production for the
four targets is given in Table~\ref{tab::table3}. The statistical
uncertainties are about 5\% for each momentum bin and nucleus. The overall
systematic uncertainties are typically 20\%, rising to 23\% for the first and
last momentum bins. The main sources of the systematic effects are related to
the background subtraction in the $K^+K^-$ invariant mass spectra
(5\,-\,10\%), the simulation of acceptance corrections
$\varepsilon_{\textrm{acc}}$ (5\,-\,10\%), the determination of the
range-telescope efficiency $\varepsilon_{\textrm{tel}}$ (10\%), and the
estimation of the integrated luminosity $L^A_{\textrm{int}}$ (12\,-\,14\%).

\begin{table*}[ht]
\caption{The measured double differential cross section
$d^2\sigma_{\phi}^{A}/(dp\,d\Omega)$ [$\mu$b/(sr\,GeV/$c$)] for $\phi$
production at small angles ($\theta_{\phi} \le 8^{\circ}$) for different
momentum bins and nuclei. The first errors are statistical and the second
systematic, which are associated with the fit quality and include the
uncertainty in the average detection efficiency $\langle \varepsilon_{\phi}
\rangle$. In addition there are overall systematic uncertainties of about
20\,-\,23\%. For the details, see the text. \label{tab::table3}}
\begin{center}
\begin{tabular}{|c|c|c|c|c|}
\hline%
$p_{\phi}$ [GeV/$c$] & C & Cu & Ag & Au\\
\hline
0.600--0.825 & $9.9\pm0.4\pm0.9$  & $26.2\pm1.3\pm2.7$ & $43.3\pm2.0\pm4.3$ & $58.0\pm2.4\pm5.2$ \\
0.825--0.950 & $13.3\pm0.6\pm0.8$ & $34.4\pm1.7\pm2.4$ & $46.6\pm2.3\pm4.0$ & $72.0\pm3.0\pm4.0$ \\
0.950--1.075 & $14.5\pm0.6\pm1.0$ & $37.3\pm1.8\pm2.6$ & $51.0\pm2.3\pm4.0$ & $76.8\pm3.0\pm4.9$ \\
1.075--1.200 & $15.3\pm0.7\pm1.5$ & $40.3\pm1.8\pm3.2$ & $55.8\pm2.7\pm4.4$ & $76.7\pm3.1\pm6.0$ \\
1.200--1.325 & $18.1\pm0.8\pm1.0$ & $40.9\pm2.1\pm2.6$ & $57.8\pm2.8\pm3.4$ & $83.5\pm3.8\pm3.4$ \\
1.325--1.600 & $18.7\pm0.7\pm0.9$ & $41.4\pm1.9\pm1.8$ & $53.1\pm2.4\pm2.2$ & $77.2\pm3.4\pm2.9$ \\
\hline
\end{tabular}
\end{center}
\end{table*}
%
%
\section{Discussion}

To interpret the data presented here, a reaction model is essential and, in
the subsequent discussion, we consider three approaches.\\
Model 1: The eikonal approximation of the Valencia group~\cite{Magas:2004}
uses the predicted $\phi$ self-energy in nuclear
medium~\cite{Cabrera:2002,Cabrera:2003} both for the one-step ($pN \to
pN{\phi}$) and two-step $\phi$ production processes, with nucleon and
$\Delta$ intermediate states.\\%
Model 2: Paryev~\cite{Paryev:2009} developed the spectral function approach
for $\phi$ production in both the primary proton-nucleon and secondary
pion-nucleon channels.\\%
Model 3: The BUU transport calculation of the Rossendorf
group~\cite{Rossendorf} accounts for a variety baryon-baryon and meson-baryon
$\phi$ production processes. In contrast to models 1 and 2, where $\phi$
absorption is governed by its width, $\Gamma_{\phi}$, model 3 describes it in
terms of an effective in-medium $\phi N$ absorption cross section
$\sigma_{\phi N}$ that can be related to the $\phi$ width $\Gamma_{\phi}$
within the low-density approximation.

One of the major differences between the three models is in the treatment of
the secondary $\phi$ production processes. In addition, in contrast to model
2, in model 3 a $\phi$ mass shift of $-22$~MeV/$c^2$ at density $\rho_0$ is
considered. This results in a modest increase of the $\phi$ production cross
section in the range of 0.6\,-\,1.6~GeV/$c$.

\begin{figure*}[ht]
  \vspace*{+0mm}
  \includegraphics[clip,width=1.6\columnwidth]{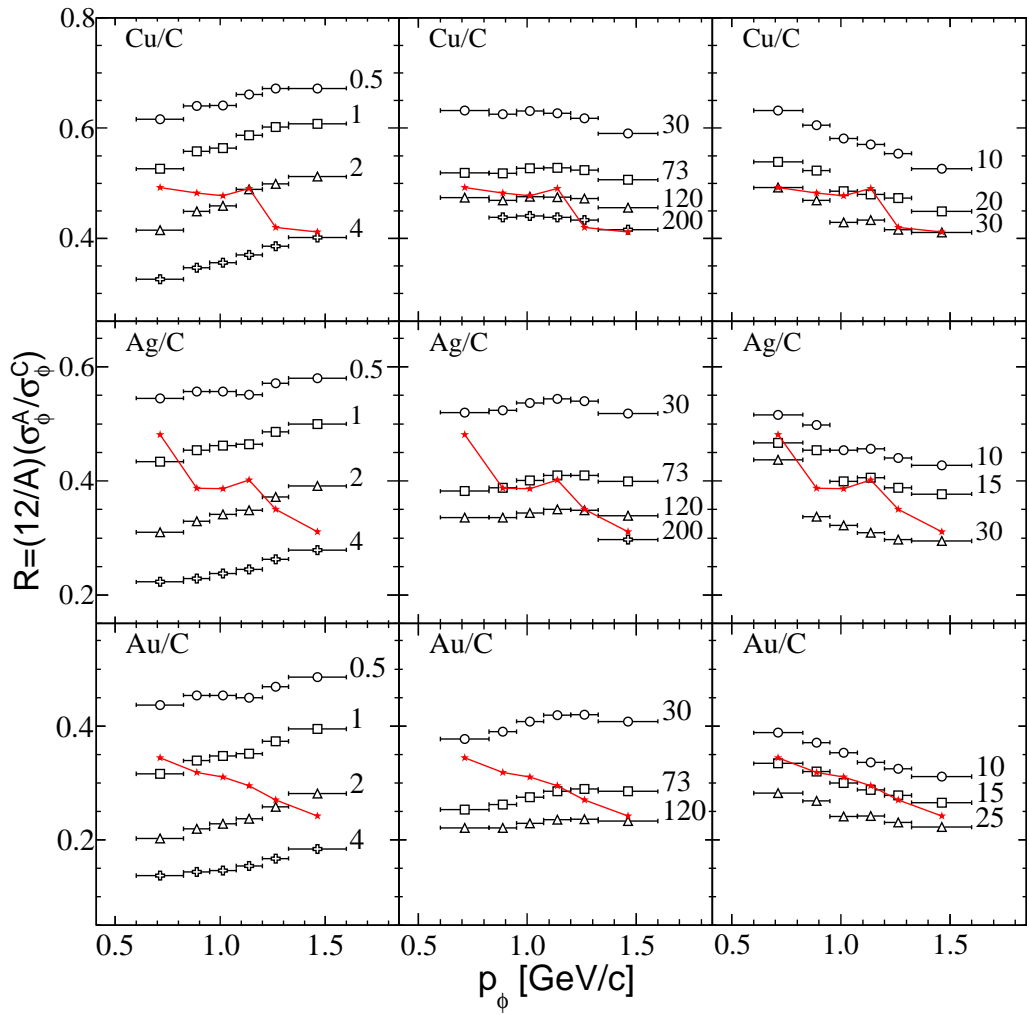}
  \vspace*{-2mm}
\caption{(Color online) The transparency ratio $R$ for the
three different nuclear combinations as a function of the
$\phi$ laboratory momentum. The experimental data, shown in
red, are connected with lines to guide the eye. The predictions
of the three theoretical approaches are shown for model 1
(left), model 2 (center), model 3 (right). For the rest of the
notations, see the text. }\label{fig:TRMod} \vspace*{-2mm}
\end{figure*}

The results of the calculations of the transparency ratio as functions of the
$\phi$ momentum are presented separately in the three columns of
Fig.~\ref{fig:TRMod} for the different models. Curve 1 of model 1 results
from using the predictions~\cite{Cabrera:2003} for the imaginary part of the
$\phi$ self-energy in nuclear matter. The other curves, corresponding to
calculations with this self-energy multiplied by factors of 0.5, 2 and 4,
demonstrate the sensitivity of $R$ to the value of the $\phi$ width.
Calculations within model 2 shown in the central panel were performed with
different values of a momentum-independent width $\Gamma_{\phi}$ of the
$\phi$ meson in its rest frame at density $\rho_0$. These values are noted
(in MeV/$c^2$) next to the curves. Calculations in model 3 presented in the
right panel were produced using the absorption cross section $\sigma_{\phi
N}$ (in mb) indicated. For comparison, the values of the experimental
transparency ratios are also shown, but without error bars.

It is seen from Fig.~\ref{fig:TRMod} that, for an almost momentum-independent
$\phi$ self-energy, model 1 predicts a steady rise of the transparency ratios
with laboratory $\phi$ momentum, which is inconsistent with the steady fall
in the data. Any momentum dependence in the results of model 2 is much more
moderate but it is clear that in neither case can the variation of the Cu/C,
Ag/C and Au/C transparency ratios be described with a single value of the
$\phi$ width. The results from model 3 are more promising in this respect,
since the steady fall in the data can be reproduced with a constant ${\phi}N$
absorption cross section of about 15\,-\,20~mb.

By comparing the calculated and measured values of the transparency ratio for
the three target combinations, it is possible to determine the weighted
average of the $\phi$ width $\Gamma_{\phi}$ in the nuclear rest frame for
density $\rho_0$ for each momentum bin. The results of applying this
procedure are shown in Fig.~\ref{fig:GamSig}(a) for both model 1 and 2. Model
3, as well as the SPring-8~\cite{Ishikawa:2004} and JLab~\cite{Wood:2010}
data, are directly sensitive to the values of the $\phi N$ absorption cross
section that are noted in Fig.~\ref{fig:GamSig}(b). The values of
$\Gamma_{\phi}$ shown in Fig.~\ref{fig:GamSig}(a) were, in these cases,
deduced in the low-density approximation, $\Gamma_{\phi} =
p_{\phi}\rho_0\sigma_{\phi N}/E_{\phi}$.

\begin{figure}[ht]
  \vspace*{+0mm}
  \includegraphics[clip,width=0.9\columnwidth]{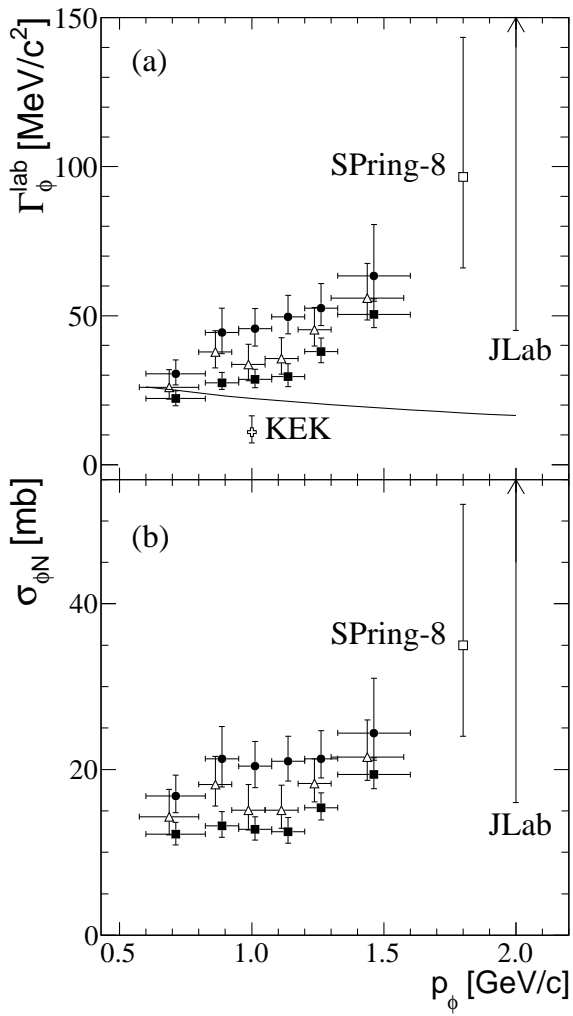}
  \vspace*{-2mm}
\caption{(a) In-medium width of the $\phi$ meson in the nuclear
rest frame at saturation density $\rho_0$ as a function of the
$\phi$ momentum. The points have been evaluated by comparing
the data of Fig.~\ref{fig:TRExp} with the results of the three
model calculations shown in Fig.~\ref{fig:TRMod} (model 1~--
full squares, model 2~-- full circles and model 3~-- open
triangles). Also shown are the results from the other experiments
noted~\cite{Muto:2005,Ishikawa:2004,Wood:2010}. The solid line
represents the $\Gamma_{\phi}$ calculated on the basis of the
predicted $\phi$ self-energy in nuclear
matter~\cite{Cabrera:2003}. (b) The $\phi N$ absorption
cross section. In the case of model 3, this is the parameter
that is directly determined from the comparison with
experiment, whereas for models 1 and 2 it is deduced from the
in-medium $\phi$ widths within the low-density approximation.
The SPring-8~\cite{Ishikawa:2004} and JLab~\cite{Wood:2010}
values of the cross section are also shown. }\label{fig:GamSig}
\vspace*{-2mm}
\end{figure}

Figure~\ref{fig:CrSec} shows the measured differential cross sections for
$\phi$ production as functions of $p_{\phi}$. The result on the light C
nucleus increases much faster with the $\phi$ momentum than for heavier
targets, and this is reflected in the variation of the transparency ratio in
Fig.~\ref{fig:TRExp}. The experimental results are compared with the
predictions of the models 2 and 3 that use the values of the $\phi$ width and
$\phi N$ absorption cross section shown in Fig.~\ref{fig:GamSig}. The
agreement of both models with the data generally improves for larger
$p_{\phi}$ though, in the highest momentum bin, the results of model 3 lie
closer to experiment. One possible reason for this is the introduction of a
greater number of $\phi$ production channels in this model. On the other
hand, both models underestimate strongly the experimental data at low
$p_{\phi}$.

The models are, of course, sensitive to the relative strength of $\phi$
production in $pp$ and $pn$ collisions~\cite{Polyanskiy:2011}. This is
experimentally uncertain and a theoretical estimate of the
ratio~\cite{Kaptari:2005} was used within the models. This corresponds to the
cross section for $\phi$ production in $pn$ collisions being about four times
larger than in $pp$ collisions at 2.83~GeV. This point is particularly
significant for the high momentum components.

An alternative way to estimate the in-medium $\phi$ width or $\sigma_{\phi
N}$ would be through a direct fit of the absolute cross sections within the
framework of either model 2 or 3. It is clear that the uncertainties in the
resulting parameters would be larger than those that use the transparency
ratio because there are then no cancelations, either theoretical or
experimental. However, calculations within these models show that, for
$p_{\phi} > 1$~GeV/$c$, the production cross sections can be described with a
$\phi$ width of about 40~MeV/$c^{2}$ at density $\rho_0$, i.e., with a $\phi
N$ absorption cross section of 15\,-\,20~mb. These values are consistent with
those shown in Fig.~\ref{fig:GamSig}. At lower momenta, the calculations in
both models underestimate the data even if one takes the free value of the
$\phi$ width or a vanishing value for the $\phi N$ absorption cross section.

The above inconsistencies suggest that some processes, whose contributions to
the $\phi$ production cross sections increase for lower $\phi$ momenta and
with the size of the nucleus, are not present in the models. The inclusion of
additional secondary production reactions, involving for example $\omega N
\to \phi N$~\cite{Sibirtsev:2006}, as well as processes where the $\phi$
slows down during its propagation through the nucleus through elastic and
inelastic collisions~\cite{Muehlich:2003}, would enhance the low-momentum
part of the $\phi$ spectrum. Unfortunately, the cross sections for such
processes are not known experimentally and so cannot be introduced reliably.

It can be argued that the transparency ratio is less sensitive to nuclear
effects and secondary production processes than the production cross section.
This may provide some justification for using the models to extract the
$\phi$ width from the experimental transparency ratio over the full momentum
range studied. In fact, model 1 allows one to deduce values of
$\Gamma_{\phi}$ from the $A$-dependence of $R$, while not making any
predictions for the $\phi$ production cross sections.

As can be seen from Fig.~\ref{fig:GamSig}, the application of the three
models yields broadly consistent results. The differences come mainly from
the divergent descriptions of the secondary $\phi$ production processes. Our
findings are not inconsistent with the KEK result, taking into account the
uncertainties in both the experiment and in the model-dependent analysis. The
observed growth of $\Gamma_{\phi}$ with $p_{\phi}$ is supported by the
SPring-8 and JLab data. Note that the values of $\Gamma_{\phi}$ extracted
within the three models agree quite well with those predicted by the Valencia
group~\cite{Cabrera:2003} at $\phi$ momenta between 0.6 and 0.825~GeV/$c$ but
deviate strongly from them at higher $p_{\phi}$, reaching a magnitude of
about 50\,-\,70~MeV/$c^2$ at $p_{\phi} \approx 1.5$\,-\,1.6~GeV/$c$. It is
worth noting that, taken together, the latest results by the
CBELSA/TAPS~\cite{Kotulla:2008} and CLAS/JLab~\cite{Wood:2010} collaborations
suggest an increase also of the in-medium $\omega$ meson width with
$p_{\omega}$.

Our data shows evidence for a momentum dependence of $\sigma_{\phi N}$ and,
as a consequence, our findings are not inconsistent with the results from
SPring-8 and JLab. The absorption cross section is between 14 and 21~mb in
the $p_{\phi}$ range of 0.6\,-\,1.6~GeV/$c$. This is also in line with the
value $\sigma_{\phi N} > 20$~mb deduced by the CLAS Collaboration from a
combined analysis of coherent and incoherent $\phi$ production from
deuterium~\cite{Qian:2009}.

\begin{figure}[ht]
\vspace*{+0mm}
\includegraphics[clip,width=1.0\columnwidth]{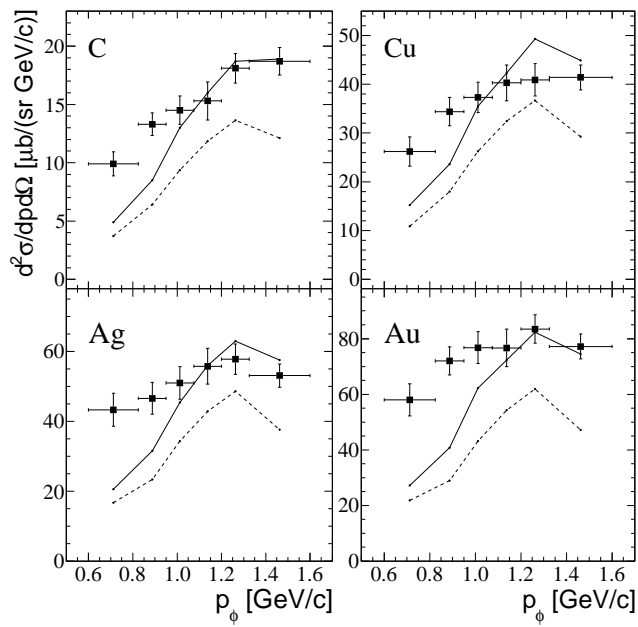}
\vspace*{-2mm} \caption{Inclusive double-differential cross
sections for $\phi$ production at small angles, $\theta_{\phi}
< 8^{\circ}$, in the collisions of 2.83~GeV protons with C, Cu,
Ag, and Au targets as functions of the $\phi$ laboratory
momentum (full squares). The errors shown are those from
Table~\ref{tab::table3} added in quadrature. The experimental
data are compared with the predictions of model 2 (dashed lines)
and model 3 (solid lines) using, respectively, the central values
of the $\phi$ width and effective $\phi N$ absorption cross section
shown in Fig.~\ref{fig:GamSig}.}\label{fig:CrSec} \vspace*{-2mm}
\end{figure}
%
%
\section{Conclusions}

The differential cross sections for the forward production of $\phi$ mesons
by 2.83~GeV protons incident on nuclear targets have been measured at the
ANKE-COSY facility. The dependence of the transparency ratio on the $\phi$
momentum was determined over the range 0.6\,-\,1.6~GeV/$c$. Values of the
$\phi$ width in nuclear matter were extracted by comparing these data with
calculations carried out within the available models. Independent of the
model used for the analysis, the results show evidence for an increase of the
$\phi$ meson width with its momentum. This was completely unexpected and
represents possibly a significant result.

Sizable excesses have been observed in the numbers of $\phi$ mesons produced
with momenta below 1~GeV/$c$. These are not reproduced by the models employed
and might suggest some enhancement in the low mass $\phi N$ systems. In order
to get a deeper insight into the momentum dependence of the $\phi$ meson
in-medium width, a better understanding of both the $\phi$ production
mechanism and its propagation through nuclear matter is crucial.

In general, $\phi$ meson production on hydrogen with elementary probes is not
completely understood at the energy of our
experiment~\cite{Mibe:2005,Dey:2011} and this should certainly be improved.
It might be interesting to note in this context that strangeness production
in closely related channels might have some influence
here~\cite{Dey:2010hh,AnefalosPereira:2009zw,Kohri:2009xe}.

%
%
\begin{acknowledgments}
We would like to dedicate this paper to our friend and colleague Vladimir
Petrovich Koptev, who died in January. Support from the members of the ANKE
Collaboration, as well as the COSY machine crew, are gratefully acknowledged.
We are particularly appreciative of the help and encouragement that we
received from Eulogio Oset. This work has been partially financed by the
BMBF, COSY FFE, DFG, and RFBR.

\end{acknowledgments}


\end{document}